# Dynamic Sensitivity Study of MEMS Capacitive Acceleration Transducer Based on Analytical Squeeze Film Damping and Mechanical Thermoelasticity Approaches

Liangrid L. Silva[1,3*]; Janderson R. Rodrigues[1,3], Angelo Passaro[1,3], Vilson R. Almeida[2,3]

[1]*Institute for Advanced Studies –Applied Physics Group, S. J. Campos – SP – Brazil*
[2]*Institute for Advanced Studies –Photonics Group, S. J. Campos – SP – Brazil*
[3]*Aeronautics Institute of Technology, São José dos Campos – SP – Brazil*
*\*liangrid@ita.br*

**Abstract:** The dynamic behavior of a capacitive micro-electro-mechanical (MEMS) accelerometer is evaluated by using a theoretical approach which makes use of a squeeze film damping (SFD) model and ideal gas approach. The study investigates the performance of the device as a function of the temperature, from 228 K to 398 K, and pressure, from 20 to 1000 Pa, observing the damping gas trapped inside de mechanical transducer. Thermoelastic properties of the silicon bulk are considered for the entire range of temperature. The damping gases considered are Air, Helium and Argon. The global behavior of the system is evaluated considering the electro-mechanical sensitivity ($S_{EM}$) as the main figure of merit in frequency domain. The results show the behavior of the main mechanism losses of SFD, as well as the dynamic sensitivity of the MEMS transducer system, and are in good agreement with experimental dynamic results behavior.

**Keywords**: electro-mechanical sensitivity; Dynamic sensitivity, micro-electro-mechanical systems (MEMS); capacitive micro-accelerometer, squeeze film damping, thermoelasticity; sensors.

## 1. Introduction

The impact of micro-electro-mechanical systems (MEMS) technology in the modern society is well known [1], [2].

This technology has been allowing the development of small, low-cost, and robust devices and systems, such as microsensors and microactuators. Nowadays, microsensors are widely used in many commercial products, such as automotive airbag systems, smartphones, tablets, and video games. Furthermore, high performance MEMS sensors are used in key areas such as defense and aerospace industries for instance, to integrate inertial measurement units (IMU's) of navigation systems applied in micro and nano satellites [3],[4],[5].

MEMS capacitive accelerometers detect accelerations by exploiting the movement of a seismic mass, resulting in a capacitance change among parallel plates suspended mechanically by clamped beams. A three-dimensional sketch of a MEMS capacitive acceleration transducer is shown in exploded view at Figure 1. In this case, the microsensor is fabricated using a bulk-micromachined process in a silicon monocrystalline substrate (Si). Such process involves a selective anisotropic wet etching, usually using potassium hydroxide (KOH) as the etching agent [1].

There are several parameters used to completely characterize this kind of device, but the main requirements are defined according to their applications. The most common figures of merit are sensitivity, dynamic bandwidth, non-linearity, resolution, etc. [6]. Sensitivity and dynamic bandwidth are the parameters explored in this paper. The sensitivity, evaluated in a static point of view, depends on materials physical properties and on geometrical parameters. The dynamic bandwidth is related to a relatively flat bandwidth and linear phase responses up to some maximum specified frequency, related to the device´s natural frequency. The evaluation of these combined parameters is assumed as a dynamic sensitivity. The performance of MEMS devices is associated with losses due to several damping effects, such as: damping of acoustic vibrations, damping due to cantilevers linking the seismic mass to the inertial frame, damping due to thermoelastic effects, intrinsic damping material, and squeeze film damping (SFD) [7], [8]. The last one is the predominant loss effect in planar MEMS structures [8]. One of the strategies used to improve the device dynamic performance is the adjustment of the atmosphere surrounding the seismic mass, in order to control the SFD effect. It is possible to set the internal atmosphere in the capacitive MEMS during the packaging process.

Numerical methods, such as the finite element (FEM), the finite volume (FVM) and the finite difference (FDM) methods, are frequently used to evaluate the multiphysical behavior of this kind of devices. Such methods demand considerably computational resources. The required computational resources increase as the complexity of the geometrical and physical models increases. When possible, simplified analytical approaches are the best way to validate first concept designs and to take the first insights on the nature of the system.  Based on analytical approximations, several squeeze film damping models, SFD, have been applied in the study of MEMS structures, such as the molecular models developed and tested by Christian R. G. [9], Newell [10], Kádár *et al.*, [11], Li *et al.*, [12], Bao *et al.*, [13] and others.  An analytical SFD model was used by Bourgeois *et al.* [14] to study a capacitive

MEMS accelerometer in nitrogen atmosphere, for different pressures. Their results show good agreement with experimental data. Recently, a new viscous damping model was proposed in Aoust *et al*. [15] to study MEMS resonators, for which additional damping sources have become important.

In this paper, we adopt a semi-analytical model to study a capacitive MEMS accelerometer. Such model takes into account the thermoelastic stiffness and linear expansion ($\alpha_L$) coefficients of anisotropic silicon bulk. In addition, an analytical damping model derived from the Reynolds equations [16], [17] is incorporated in the model in order to study dynamic characteristics of a MEMS capacitive accelerometer. Such approach takes into account the inertial effects of the air, argon and helium gasses, assumed as ideal gasses. The simulation model was compared to experimental measurements. The main figure of merit adopted is the electromechanical sensitivity (SEM) in the frequency response perspective (dynamic sensitivity) considering the effect of pressure and of the temperature of the gas on the damping loss mechanisms in such devices.

## 2. Capacitive Acceleration Transducer Model

The capacitive transducer illustrated in Fig. 1 allows detecting an external force by the difference of capacitance between the fixed conductor plates and the seismic mass, displaced from the equilibrium by this force. This difference is converted to an electrical voltage signal [5]. The electromechanical transducer can be modeled as a mass-spring-damper system, as illustrated in Figure 2.

The complete system can be modeled by a non-linear differential equation with time-dependent coefficients, given by [4]:

$$\frac{d^2 z_v}{dt^2} = \frac{d^2 z}{dt^2} + \frac{\eta_f W_m^4 \beta(\Gamma)}{Q_{pr} M_{eff}} \left[\frac{1}{(h_0 - z)^3} + \frac{1}{(h_0 + z)^3}\right] \frac{dz}{dt} + \frac{k_{eq}}{M_{eff}} + \frac{-\varepsilon_r \varepsilon_o W_m^2 V_s^2}{2 M_{eff}} \left[\frac{1}{(h_0 - z)^2} - \frac{1}{(h_0 + z)^2}\right] \quad (1)$$

where $M_{eff}$ is the effective mass of the seismic mass, $W_m$ is the width of the seismic mass, $k_{eq}$ is the equivalent stiffness constant of the beams set, $V_s$ is the excitation voltage signal, $\eta_f$ is the viscous fluid coefficient, $\varepsilon_r$ is the relative electrical permissivity of the gas, $\varepsilon_0$ is the electrical permittivity of vacuum, $\beta(\Gamma)$ is the geometric correction factor, $Q_{pr}$ is the relative rate flux coefficient, $h_0$ is the distance between fixed and mobile plates, $z$ is the displacement coordinate, and $z_v$ is the displacement related with the inertial frame.

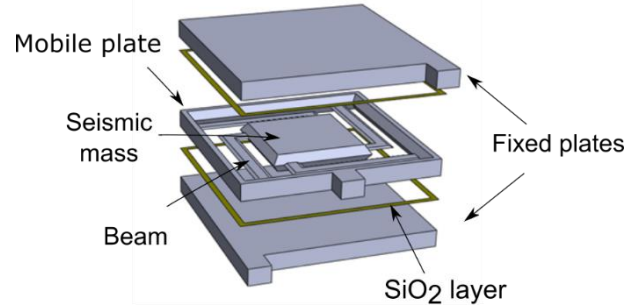

*Figure 1 – Exploded view of a MEMS capacitive acceleration transducer.*

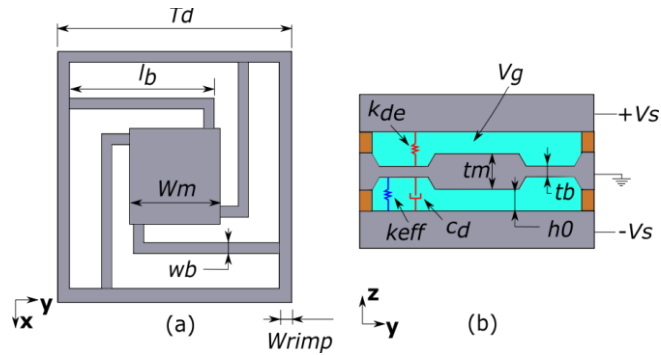

Figure 2 – Schematic illustration of the geometric structure of the MEMS capacitive acceleration transducer. (a) front view of the seismic mass (mobile plate). (b) side view showing the internal parts of the transducer. The variables are presented in

TABLE 4 – COMPARISON OF PARAMETERS OF CAPACITIVE MEMS ACCELEROMETER DEVICE, BETWEEN EXTREME LIMITS RANGE TEMPERATURE SIMULATIONS WITH REFERENCE TEMPERATURE PARAMETERS VALUES. STATIC CONDITION AT 1ATM PRESSURE IN AIR GAS MEDIUM.

| Description | Nominal Values at 293 K | Values at 228 K [%] | Values at 398 K[%] |
|---|---|---|---|
| Length of the seismic mass ($Wm = a$) | 2.0E+03 μm | < 1.0E-03 % | < 7.0E-03 % |
| Width of the seismic mass ($Wm = b$) | 2.0E+03 μm | < 1.0E-03 % | < 7.0E-03 % |
| Width of the beam ($wb$) | 1.77E+02 μm | < 1.0E-03 % | < 1.0E-03 % |
| Thickness of the beam ($tb$) | 5.5E+01 μm | < 1.0E-03 % | < 1.0E-03 % |
| Thickness of the seismic mass ($tm$) | 3.8E+02 μm | < 1.0E-03 % | < 2.0E-03 % |
| Length of the beam ($lb$) | 2.797E+03 μm | < 1.0E-03 % | < 2.0E-03 % |
| Width of rim ($wrimp$) | 1.5E+02 μm | < 1.0E-03 % | < 2.0E-03 % |
| Total length of transducer ($Td$) | 4.38E+03 μm | < 1.0E-03 % | < 1.0E-03 % |
| Effective stiffness constant ($keff$) | 1.113E+03 Nm | +4.1E-01 % | -6.7E-01 % |
| Effective mass ($meff$) | 4.21585E-06 kg | -5.4E-04 % | +9.5E-04 % |
| Volume of the gas ($Vg$) | 4.301E+03 cm3 | < 1.0E-12 % | < 1.0E-12 % |
| Gap between fixed electrodes and the seismic mass ($h0$) | 2.0E+00 μm | < 1.0E-03 % | < 1.0E-03 % |

and TABLE 5.

For small displacements of the seismic mass, the equation can be linearized, for instance, by using a Taylor series around the operating point $z = 0$ [18], [19]. Making use of linear systems analysis techniques, we can write the transfer function of the system, relating the $z$ axis displacement of the seismic mass and the applied acceleration $a_z$, taking into account the inertial effects of squeeze film damping gas, as follow [5]:

$$H_{EM} = \frac{Z}{A_z} = \frac{k_{em}}{K_2 s^2 + K_1 s + K_0} \quad (2)$$

where $k_{em}$ is the electromechanical constant, $K_2 = 1$,

$$K_1 = 2c_d M_{eff}^{-1}(k_{eq} M_{eff})^{\frac{-1}{2}}(k_{eff} + 2k_{de})^{\frac{1}{2}}, \quad (3)$$

$$K_0 = k_{eff} + 2k_{de} M_{eff}^{-1}, \quad (4)$$

$c_d$ is the viscous gas damping coefficient, and $k_{de}$ is the elastic gas damping coefficient [17]. In order to take into account the damping force due to the gas film located between the faces of the seismic mass and the fixed plates, a gas film lubrication approach derived from the modified Reynolds equation [16], [20] and based on the continuous theory of matter is employed [17]:

$$\frac{\partial}{\partial x}\left(\frac{\rho_f h_0^3}{12\eta_f} Q_{fr} \frac{\partial p}{\partial x}\right) + \frac{\partial}{\partial y}\left(\frac{\rho_f h_0^3}{12\eta_f} Q_{fr} \frac{\partial p}{\partial y}\right) = p_f \frac{\partial h_0 \rho_f}{\partial t} \quad (5)$$

where $\rho_f$ is the fluid density, $p_f$ is the fluid pressure and $\eta_f$ is the fluid viscosity. Such approaches are named Squeeze Film Damping models (SFD models). Equations from (1) to (5) allow an improvement in the representation of the dynamic behavior of the actuator by including gas film inertial effects. The inertial effects alter the natural frequency and the peak amplitude as a function of applied pressure [5]. The effect of the elastic gas damping coefficient is neglected for squeeze number smaller than 0.2 and under this conditions the predominant damping loss mechanism is the viscous damping [8], [13]. Assuming that the resultant force oscillates harmonically and moves the seismic mass, the relation between the damping coefficients is expressed by the damping force, $F_d$, on the plates due to the squeeze effect and it is given by following equation:

$$F_d = c_d + k_{de} = \Re[\Phi]Zs - \omega Im[\Phi]Z \quad (6)$$

The first and second terms in the right side of eq. (6) correspond to the viscous gas damping coefficient and the elastic gas damping coefficient, respectively. where $\omega$ is the frequency, $\Re[.]$ and $Im[.]$ denotes the real and imaginary parts of the complex number. The force of fluid per unit of speed $\Phi$ is expressed in terms of the infinite series summation over odd indices $m$ and $n$ containing the terms given by [17], [21]:

$$\Phi = \sum_{m,odd}^{M} \sum_{n,odd}^{N} \frac{1}{Q_{fr} G_{mn} + j\omega D_{mn}} \quad (7)$$

Considering that the fixed and mobile plates (seismic mass) are rectangular, the geometric terms $G_{mn}$ and $D_{mn}$ are calculated respectively as:

$$G_{mn} = \frac{(mn)^2 \pi^6 h_0^3}{768 \eta_f ab}\left(\frac{m^2}{a^2} + \frac{n^2}{b^2}\right) \tag{8}$$

$$D_{mn} = \frac{(mn)^2 \pi^4 h_0}{64 \eta_f abpn_\gamma} \tag{9}$$

where, $a$ and $b$ are the lengths of the edges of the seismic mass, $n_\gamma$ is the isothermal process coefficient, which depends on the heat conduction and the temperature assumed as boundary conditions during the linearization of the modified Reynolds equation in frequency-domain [17]. The relative flow rate model, $Q_{fr}$, which takes into account high squeeze effect number assuming inertial gas effects can be defined as follows [17]:

$$Q_{fr_K} = \sum_{k,odd}^{K} \frac{1 + 6K_s}{\frac{k^4 \pi^4}{96} + \frac{j\omega k^2 \pi^2 \rho_f h_0^2 (1 + 10K_s + 30K_s^2)}{96\eta_f(1 + 6K_s)}} \tag{10}$$

where $k=1, 3, 5,.. K$ and $K_s$ is the rarefaction coefficient which is given as follows:

$$K_s = \sigma_p K_n \tag{11}$$

The *slip-flow* coefficient $\sigma_p$ includes the effect of kinetics velocity variation of gas molecules due to thermal energy near the surface and it is given by [22]:

$$\sigma_p = \frac{2 - \alpha_v}{\alpha_v}[1.016 - 0.1211(1 - \alpha_v)] \tag{12}$$

where, $\alpha_v$ is the coefficient that expresses the diffusion reflected fraction of gas molecules, also known as *tangential momentum accommodation coefficient* (TMAC) [22]. The Knudsen number is given as follows:

$$K_n = \frac{\lambda_{lc}}{h_o} \tag{13}$$

$$\lambda_{lc} = \frac{K_B T}{\sqrt{2}\sigma_{ch} p} \tag{14}$$

where, $\lambda_{lc}$ is the mean free path of gas assuming elastic collisions [17], $K_B$ is the Boltzmann constant, $T$ is the temperature, and $\sigma_{ch}$ is the effective collision area of gas. The equations which evaluate the amplitude and phase of dynamic response of the MEMS capacitive transducer are given as follow [8]:

$$A_0 = \frac{f_0}{Meff}\sqrt{\frac{1}{(\omega_a^2 - \omega^2)^2 + c_d^2 \omega^2 / Meff^2}} \tag{15}$$

$$\theta_d = -arctan\left(\frac{c_d \omega}{Meff(\omega_a^2 - \omega^2)}\right) \tag{16}$$

where, $f_o$ is the amplitude of external force; $\omega_a$: modified frequency, which is a function of the angular frequency with damping effects, and can be obtained by:

$$\omega_a = \sqrt{\frac{k_{eff} + k_{de}}{Meff}} \tag{17}$$

The damping factor $\xi_m$ and the angular natural frequency $\omega_{rm}$ equations are, respectively:

$$\xi_m = \frac{c_d}{2(k_{eq}M_{eff})^{\frac{1}{2}}} \tag{18}$$

$$\omega_{rm} = \left(\frac{k_{eff}}{M_{eff}}\right)^{\frac{1}{2}} \tag{19}$$

where, $k_{eff}$ is the linearized effective mechanical stiffness coefficient given by [19]:

$$k_{eff} = \frac{48EI}{l_b^3} - \frac{2\varepsilon_r\varepsilon_o W_m^2 V_s^2}{h_0^3}, \tag{20}$$

$E$ is the Young's Modulus, $I$ is the transversal moment of inertia of beams, and $l_b$ is the length of the beams. The left hand term in eq. (20) which includes the electrostatic spring softening contribution [23]. The accelerometers seismic mass with the geometric shape described previously, can be obtained in function of the mask design by [4]:

$$M_s = \rho_{si}\left(W_m^2 t_m + \frac{\sqrt[2]{2}}{2} W_m t_m^2 + \frac{t_m^3}{6}\right) \tag{21}$$

where $\rho_{Si}$ is the silicon density. The effective mass of beam $m_{eff}$ is obtained using the Rayleigh principle and can be calculated by:

$$M_{eff} = M_s + n\frac{13}{35}m_b, \tag{22}$$

where $m_b$ is the mass of the beam, $n$ is the number of beams; in this case n is equal to 2, i.e., there are two clamped-clamped beams. The mechanical static sensitivity of transducer can be expressed as [5]:

$$S_M = \left.\frac{\partial \delta_z}{\partial a_z}\right|_{\omega=0} \rightarrow S_M = \frac{1}{\omega_{rm}^2} \tag{23}$$

The electrical sensitivity $S_E$ can be expressed as:

$$S_E = \frac{2\varepsilon_r\varepsilon_o W_m^2}{(h_o^2 - z^2)} \tag{24}$$

Finally, the electro-mechanical sensitivity is defined as:

$$S_{EM} = S_E S_M \tag{25}$$

The mechanic behavior of detection structure in capacitive MEMS accelerometer transducer is basically dependent of geometric configuration and mechanical material properties. The crystal silicon (Si) used has (100) orientation and [110] flat direction. The anisotropic mechanical Si properties are stiffness $C_{ijkl}$ and compliance $S_{ijkl}$, both based on the Hooke's law and normally these terms are expressed with a fouth rank tensor [24]. Through the unit cell symmetry of crystal Si representation it is common to show the $S$ and $C$ in the orthotropic context and consequently present the properties according the specific axis of interest using a reduction notation rank tensor. The $S$ and $C$ crystal Si coefficients values used in our study for [100] orientation were based on Hall [25] and Brantley works [26]. Therefore, the elastic properties are expressed by Young's modulus $E$ and the Poisson's ratio $v$, and them can be easily found using paralel and ortonormal angular projections by the $x$, $y$ and $z$ axis directions of the Silicon crystal [26]. In addition, the stiffness tensor can then be simply rotated in the orientation of interest [27].

For a crystal Si material, the linear thermal expansion coefficient has the same behavior for all crystallographic directions [24], [28]. The thermal effect contribution in the Si stiffness coefficients could be expressed surrounding a reference temperature due the expression [14], [29]:

$$C_{T_{ij}} = C_{T0_{ij}}\left[1 + \sum_{n\geq 1}\left(TCE_{C_{ij}}\right)_n (T-T_0)^n\right] \tag{26}$$

where $T_0$ is the reference temperature, $C_{T0ij}$ is the stiffness coefficient at reference temperature, $TCE_{cij}$ is the thermal coefficient of elasticity of the considered elastic constant. The first order temperature coefficients of the elastic constants are presented in TABLE 1 [14], [28].

TABLE 1 – TEMPERATURE COEFFICIENTS OF THE ELASTIC CONSTANTS FOR P-TYPE($B$) AND N-TYPE ($P$) SUBSTRATES.

| TCE | $B$ (4 Ω.cm) | $P$ (0.05 Ω.cm) |
|---|---|---|
| | First order [ × 10$^{-6}$/K] | |
| $TCE_{S11}$ | 64.73 | 63.60 |
| $TCE_{S12}$ | 51.48 | 45.79 |
| $TCE_{S44}$ | 60.14 | 57.96 |
| $TCE_{C11}$ | -73.25 | -74.87 |
| $TCE_{C12}$ | -91.59 | -99.46 |
| $TCE_{C44}$ | -60.14 | -57.96 |

The linear thermal expansion coefficient $\alpha_L$ (10$^{-6}$ K$^{-1}$) for temperatures ranging from 120 K to 1500 K is expressed by [30]:

$$\alpha_L = 3{,}725 \left(1 - e^{-5{,}88 \times 10^{-3}(T-124)}\right) + 5{,}548 \times 10^{-4} T \tag{27}$$

The thermofluid considerations were based on the ideal gases law. The dynamic viscosity was given by Sutherland formula [31]:

$$\eta_f = \eta_0 \left(\frac{T}{T_0}\right)^{\frac{3}{2}} \left(\frac{T_0 + R}{T + R}\right) \tag{28}$$

where $R$ is the Sutherland constant of the gases, which is related with the potential energy of molecular gas due the mutual attraction interaction [31], and $\eta_0$ is the dynamic viscosity at reference temperature [32]. Such parameters and additional gas coefficients are shown in TABLE 2 [33].

## 3. Results and Discussion

The results are organized in four subsections. The first one presents the SFD gas model validation. The second one discusses the temperature dependency of geometric parameters, and their implications in some physical quantities. Next one shows the dynamic response of the microaccelerometer device and highlight the main loss mechanisms related with SFD effect. The last one reveals the dynamic $S_{EM}$ and its implications in transducer performance. In order to promote such simulations appointments two main considerations were made: First, the system is closed and it is in thermodynamic equilibrium. Consequently, the isothermal process coefficient ($n_\gamma$) was assumed as a constant. In addition, the electric permittivity of all gases considered in this study are assumed as independent of temperature, due the slightly variations at low pressures [34].

TABLE 2 - GASES COEFFICIENTS

| Gas | Sutherland constant [K] | Ref. temp. [K] | $\lambda_{lc}$ at 300 K, 1 atm. [nm] | $\eta_0$ at 300 K [µPa.s] | $\rho_f$ at 273 K, 1 atm. [kg/m$^3$] | $\varepsilon_r$ at 293 K 1 atm |
|---|---|---|---|---|---|---|
| Air | 120 | 291.15 | 68.7 | 18.27 | 1.184 | 1.0005364 |
| Helium | 79.4 | 273 | 198 | 19 | 0.1786 | 1.0000650 |
| Argon | 133 | 298 | 72.7 | 22.6 | 1.784 | 1.0005172 |

However, the electric permittivity of the gases have dependency with pressure and temperature [34] [35]. Silicon oxide layers are used to isolate electrically the doped-silicon plates that compound the capacitive transducer (see Figure 1). The thermal expansion coefficient of the oxide layers was assumed constant and equal to 0,24x10$^{-6}$ K for the entire range of temperature [36].

### 3.1. Model validation

In order to compare the gas squeeze film damping model which taking account inertial effects with real data measurements was performed an experimental frequency response tests. Experimental data were obtained from tests [37]. A brief description of the test is described below. The MEMS capacitive accelerometer electrostatic test of frequency response was simulated in a vacuum chamber, where different encapsulation pressures were simulated, namely: $p \in \{1000, 500, 100 \text{ and } 50 \text{ Pa}\}$. An external pick-off electronic circuit was developed to reading and interrogate the MEMS accelerometer. The experimental apparatus is illustrated in Figure 3.

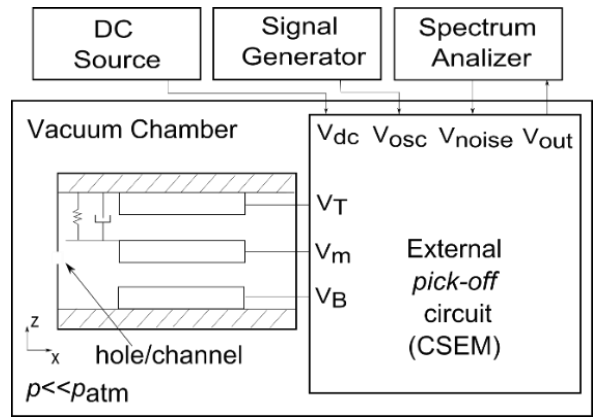

Figure 3 – Illustration of the experimental arrangement for the electrostatic test frequency response.

The theoretical model (eq. (2)) was adjusted in amplitude, offsetting the gain of the external electronic circuit [37]. For SFD model validation purposes only the pressure of 500 Pa using air gas atmosphere was plotted, in comparison with two distinct analytical simulations of SFD models (see Figure4-a). One of them capturing the inertial effects [17][5] and another one not assuming inertial effects conditions [38][4].

### 3.2. Temperature and geometric parameters

The simulations results of thermoelasticy and thermal expansion effects in the mechanical structure of the MEMS transducer, relating the main geometrical parameters and some of the main physical variables, are summarized in for two temperatures: 228 K and 398 K. This range correspond to the limits of the operational range of several standard electronic devices. The obtained values are given in terms of the variation respect the nominal parameter values at 293K. The complementary input data parameters used in this study are presented in TABLE 5. Basically, the difference lengths for almost of geometric parameters is below $1 \times 10^{-3}$ percent for 228 K, which leads a decrease in geometric lengths. At temperature of 398 K occurs an increase in these parameters and such variation is less than $7 \times 10^{-3}$ percent. The volume of gas $Vg$, is virtually constant for both temperatures. The calculated changes was less than twelve magnitude order in percentage. In view of this fact, the parameter $Vg$ was assumed as a constant and the ideal gases law behavior lead us to consider that the molar mass of all gases as constants and thereby for simplifications purposes we assume the densities of gases according the TABLE 2.

TABLE *4 – COMPARISON OF PARAMETERS OF CAPACITIVE MEMS ACCELEROMETER DEVICE, BETWEEN EXTREME LIMITS RANGE TEMPERATURE SIMULATIONS WITH REFERENCE TEMPERATURE PARAMETERS VALUES. STATIC CONDITION AT 1ATM PRESSURE IN AIR GAS MEDIUM.*

| Description | Nominal Values at 293 K | Values at 228 K [%] | Values at 398 K[%] |
|---|---|---|---|
| Length of the seismic mass ($Wm = a$) | 2.0E+03 μm | < 1.0E-03 % | < 7.0E-03 % |
| Width of the seismic mass ($Wm = b$) | 2.0E+03 μm | < 1.0E-03 % | < 7.0E-03 % |
| Width of the beam ($wb$) | 1.77E+02 μm | < 1.0E-03 % | < 1.0E-03 % |
| Thickness of the beam ($tb$) | 5.5E+01 μm | < 1.0E-03 % | < 1.0E-03 % |
| Thickness of the seismic mass ($tm$) | 3.8E+02 μm | < 1.0E-03 % | < 2.0E-03 % |
| Length of the beam ($lb$) | 2.797E+03 μm | < 1.0E-03 % | < 2.0E-03 % |
| Width of rim ($wrimp$) | 1.5E+02 μm | < 1.0E-03 % | < 2.0E-03 % |
| Total length of transducer ($Td$) | 4.38E+03 μm | < 1.0E-03 % | < 1.0E-03 % |
| Effective stiffness constant ($k_{eff}$) | 1.113E+03 Nm | +4.1E-01 % | -6.7E-01 % |
| Effective mass ($m_{eff}$) | 4.21585E-06 kg | -5.4E-04 % | +9.5E-04 % |
| Volume of the gas ($Vg$) | 4.301E+03 cm3 | < 1.0E-12 % | < 1.0E-12 % |
| Gap between fixed electrodes and the seismic mass ($h0$) | 2.0E+00 μm | < 1.0E-03 % | < 1.0E-03 % |

The effective stiffness constant $k_{eff}$, presented in eq. (20), is related with electrical and mechanical stiffness coefficients of the MEMS transducer [4]. Therefore, it is embeds the thermoelastic mechanical effects in the MEMS structure. At temperature of 228 K, $k_{eff}$ presents approximately $+4.1 \times 10^{-1}$ percent in comparison with temperature reference value, i. e. a structural behavior with higher mechanical rigidity. On the other hand, for 398 K the stiffness of structure is nearly $-6.1 \times 10^{-1}$ percent below the reference value. This implies in a structural behavior with higher mechanical elasticity. The effective mass $m_{eff}$, showed in eq.(22), presents differences around $10^{-4}$ percent in magnitude order for both temperature limits. Hence, the small variation of $m_{eff}$ values indicates a low contribution of such parameter in the global system behavior presupposing the studied temperature range.

The magnitude displacement frequency plots from air, argon and helium gases to a fixed pressure ($p = 50$ Pa) from $T \in \{228, 293$ and $398\}$ K, are shown in Figure 4 (b, c, d), respectively, assuming the value of -1 m/s$^2$ input step. The main contribution in magnitudes differences, at static condition observed for all gases, were resulted from changes, in stiffness and compliance coefficients as a result of temperature dependence at silicon structure material. In addition, air and helium gases revealed the most similar behavior when the frequency raise in comparison with argon gas. Complementary, it was observed that the dynamic behavior of argon, is more sensitive at temperature variations near the frequency of 500 Hz, compared with air and helium gases closed to the same frequency. Clearly, the linearity frequency range was affected by the gases characteristics.

TABLE 3 – COMPARISON OF VISCOUS GAS DAMPING COEFFICIENT (cd), BETWEEN EXTREME LIMITS RANGE TEMPERATURE SIMULATIONS AND REFERENCE TEMPERATURE PARAMETERS VALUES. STATIC CONDITION AT 1atm PRESSURE.

| Gas | Nominal Values at 293 K | Values at 228 K | Values at 398 K |
|---|---|---|---|
| Air | 2.39E+01 Ns/m | -1.84E+01 % | 2.63E+01 % |
| Argon | 3,84E+01 Ns/m | -1.89E+01 % | 2.71E+01 % |
| Helium | 2,57E+01 Ns/m | -1.68E+01 % | 2.36E+01 % |

The viscous gas damping coefficients $c_d$, related with eq.(6), for all studied gases maintaining the same pressure and temperature setup, are showed in

TABLE 3. As observed, Argon gas exhibited the greater values of $c_d$ for all temperatures even as, the major percent differences than the other gases. In addition, the air gas presented the smallest nominal result values compared with Helium, but the last one, presents the smallest percent values compared with the others. The elasticity gas damping coefficient $k_{de}$, which was derived from eq. (6), could be negligible in static conditions and high pressures (1 atm), because at this case, the dominant damping effect is the viscous gas damping [5], [8]. In consequence, the dominant loss mechanism in SFD approaches is the viscous gas damping.

*TABLE 4 – COMPARISON OF PARAMETERS OF CAPACITIVE MEMS ACCELEROMETER DEVICE, BETWEEN EXTREME LIMITS RANGE TEMPERATURE SIMULATIONS WITH REFERENCE TEMPERATURE PARAMETERS VALUES. STATIC CONDITION AT 1ATM PRESSURE IN AIR GAS MEDIUM.*

| Description | Nominal Values at 293 K | Values at 228 K [%] | Values at 398 K [%] |
|---|---|---|---|
| Length of the seismic mass ($W_m = a$) | 2.0E+03 μm | < 1.0E-03 % | < 7.0E-03 % |
| Width of the seismic mass ($W_m = b$) | 2.0E+03 μm | < 1.0E-03 % | < 7.0E-03 % |
| Width of the beam ($w_b$) | 1.77E+02 μm | < 1.0E-03 % | < 1.0E-03 % |
| Thickness of the beam ($t_b$) | 5.5E+01 μm | < 1.0E-03 % | < 1.0E-03 % |
| Thickness of the seismic mass ($t_m$) | 3.8E+02 μm | < 1.0E-03 % | < 2.0E-03 % |
| Length of the beam ($l_b$) | 2.797E+03 μm | < 1.0E-03 % | < 2.0E-03 % |
| Width of rim (*wrimp*) | 1.5E+02 μm | < 1.0E-03 % | < 2.0E-03 % |
| Total length of transducer (*Td*) | 4.38E+03 μm | < 1.0E-03 % | < 1.0E-03 % |
| Effective stiffness constant ($k_{eff}$) | 1.113E+03 Nm | +4.1E-01 % | -6.7E-01 % |
| Effective mass ($m_{eff}$) | 4.21585E-06 kg | -5.4E-04 % | +9.5E-04 % |
| Volume of the gas (*Vg*) | 4.301E+03 cm$^3$ | < 1.0E-12 % | < 1.0E-12 % |
| Gap between fixed electrodes and the seismic mass ($h_0$) | 2.0E+00 μm | < 1.0E-03 % | < 1.0E-03 % |

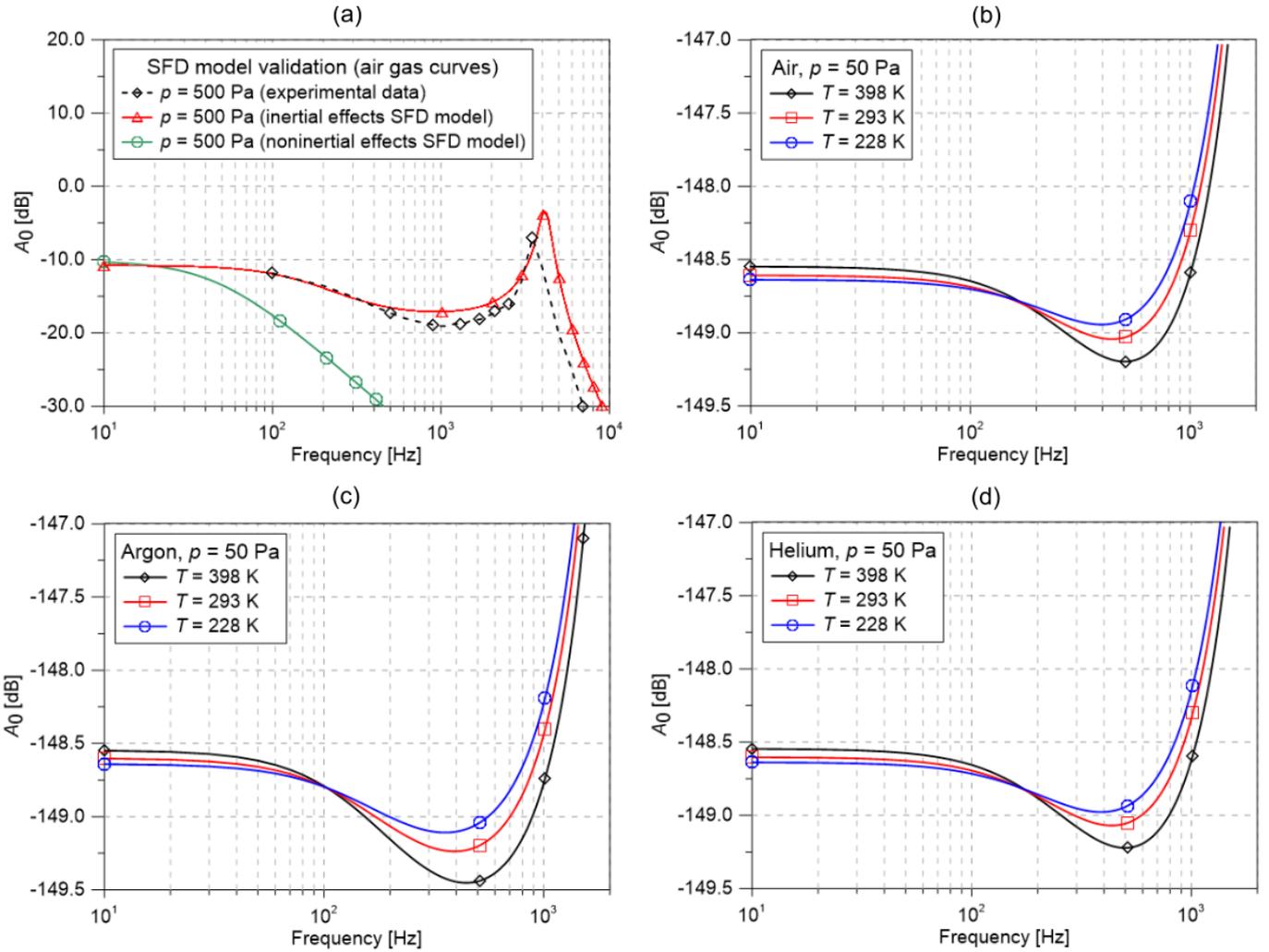

Figure 4 – (a) Squeeze film damping (SFD) gas model with inertial effects validation. The magnitude step response curves for the experimental data and two theoretical curves based on SFD gas models from pressure of 500 Pa in air atmosphere simulating the packaging conditions of MEMS capacitive acceleration transducer; (b,c,d) magnitude step responses simulation for -1 g input, using atmosphere pressure of 50 Pa in packaging for temperatures 228 K, 293 K and 398 K. (b) Air gas, (c) Argon gas and (d) Helium gas

### 3.3. Dynamic with Gas Squeeze Film Damping Assumptions

Next, for example purposes, we present the damping factor (Figure 5), the viscous gas damping coefficient (Figure 6), the elastic gas damping coefficient (Figure 7), the magnitude displacement response (Figure 8) and the phase frequency response (Figure 9) plots. All of these simulations results were obtained using the following conditions: Air gas atmosphere packaging, -1 g acceleration step response; pressure sweep range from 1000 Pa to 20 Pa; and temperature at 293 K.

We observed the improvement in linear dynamic characteristic of device occurs due to drop in pressure surrounding the seismic mass structure, which such behavior is well captured by the magnitude and phase responses and it is agreement with the physical nature of planar MEMS structures topology [17], [39]. In addition, when the pressure is reduced, the peak frequency is increased in amplitude and moves from a region of higher to lower frequencies, approaching the frequency peak of the natural mechanical resonance, of the seismic mass structure. This previously described behavior is represented by the modified frequency that was exposed in eq. (17). Also, the constant plateau at static response is observed at amplitude responses. This clearly occurs because the temperature was fixed in only one value, for examples purposes.

TABLE 5 – COMPLEMENTARY INPUT PARAMETERS OF THE STUDY.

| Description | Values |
|---|---|
| Geometric correction factor ($B_f$) | 0.4217 |
| Monocrystalline silicon density ($\rho_{si}$) | 2,330 $K_g/m^3$ |
| TMAC ($\alpha_v$) | 1.0 |
| Boltzman constant ($K_B$) | 1.38 x $10^{-23}$ J/K |
| Excitation voltage signal ($V_s$) | 5 V |
| Gravitational acceleration (g) | 9.806 $m/s^2$ |
| Isothermal process coefficient ($n_\gamma$) | 1.0 |
| Electromechanical constant ($k_{em}$) | 1.0 |

The result behavior of the damping factor coefficient are shown in Figure 5 and it is associated with eq. (18). It exhibits, for high pressures and low frequencies, the increase of parameter value. When de pressure inside the packaging decreases, the damping factor effect became less significant. The same behavior was observed, when the frequency increase. In fact, the damping factor and the viscous gas damping coefficient (Figure 6) quantifies the same behavior. The first it is a common figure of merit used in dynamic system representation, and the other is provided by SFD approaches. In fact, the viscous gas damping is the dominant loss mechanism effect in low operation frequencies at such devices. The elastic gas damping simulation was exhibited in Figure 7. At low frequencies, this coefficient can be negligible. When the frequency increase, consequently the elastic gas damping factor increase and it became dominate damping loss element in SFD effect. However, this effect can be mitigated using low pressure packaging conditions. Such parameter it is one of the main gas physical mechanisms responsible for linearity loss in the system dynamics involving gas pressure packaging in planar MEMS devices [5].

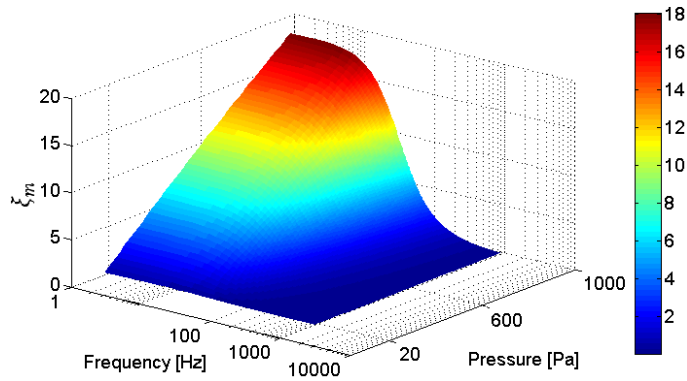

Figure 5 – Damping factor coefficient simulation of the capacitive MEMS accelerometer transducer for Air gas atmosphere packaging with pressures varying from 1000 Pa to 20 Pa.

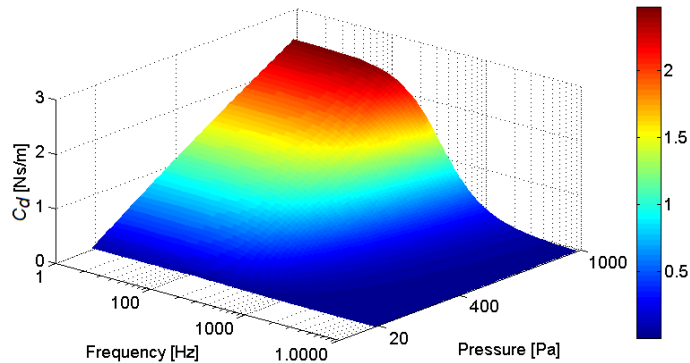

Figure 6 – Viscous gas damping coefficient simulation of the capacitive MEMS accelerometer transducer for Air gas atmosphere packaging with pressures varying from 1000 Pa to 20 Pa.

In practice, at low frequencies, the viscous gas damping forces dominates the damping mechanism, because under these conditions, the gas has enough time to escape through the side of the boards of the seismic mass structure [8]. In the opposite case, i.e. at high frequencies, the elastic gas force, of the gas increases and starts to exert the dominant effect in the system dynamics. When the tensile strength becomes dominant to the point of not giving enough time to the fluid film to escape, trapping it between the plates, the phenomenon occurs that gives rise to compression effects known as squeeze effects [8]. Consequently, when the SFD effect are minimized, the linearity of the system response is improved, and the dominant damping behavior becomes to related with the mechanical damping mechanism of the MEMS transducer structure, which is limited by the loss energy due the mechanical resonance. It is evident that, the suitable control of pressure atmosphere packaging promote the reduction of undesirable effects of SFD in the capacitive MEMS acceleration transducer.

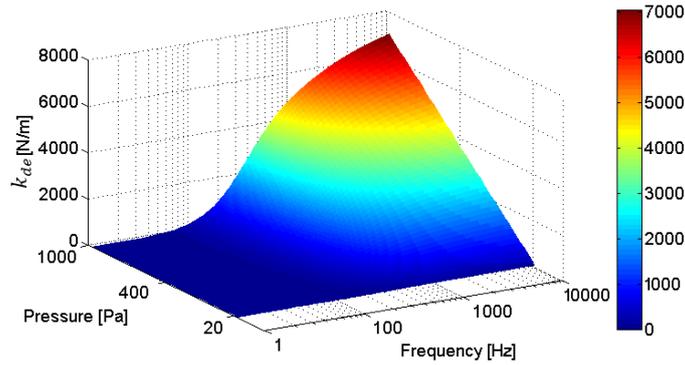

Figure 7 – Elastic gas damping coefficient simulation of the capacitive MEMS accelerometer transducer for Air gas atmosphere packaging with pressures varying from 1000 Pa to 20 Pa.

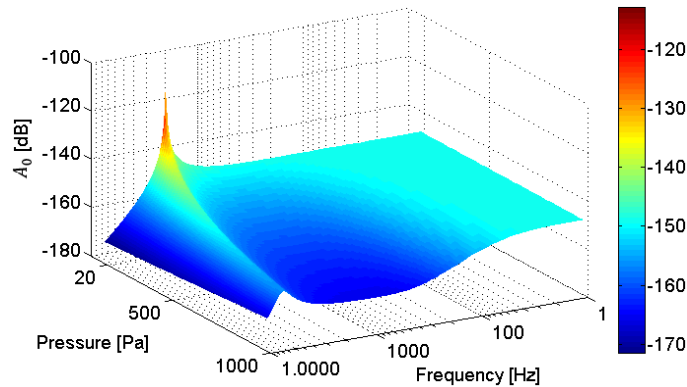

Figure 8 – Magnitude displacement step response for -1 g input in the capacitive MEMS accelerometer transducer for air gas atmosphere packaging with pressures varying from 1000 Pa to 20 Pa.

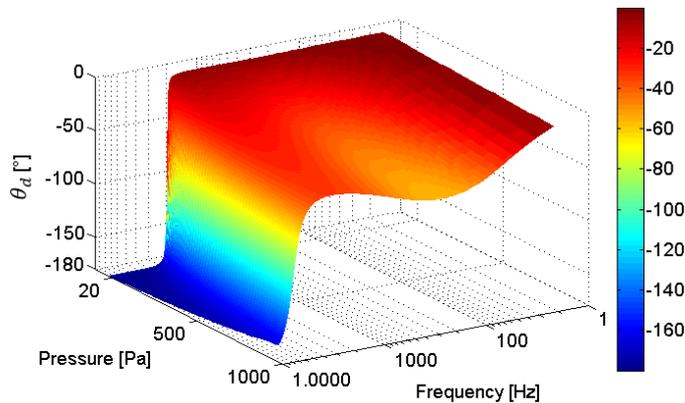

Figure 9 – Phase step response for -1 g input in the capacitive MEMS accelerometer transducer for air gas atmosphere packaging with pressures varying from 1000 Pa to 20 Pa.

### 3.4. Electro-mechanical Sensitivity Results

The electro-mechanical sensibility ($S_{EM}$) simulation results were presented, using two distinguish pressures setup (200 Pa and 20 Pa) to the temperature range from 228 K to 398 K and the frequency range from 0 to 1000 Hz, for air (Figure 10, Figure 11), argon (Figure 12, Figure 13) and helium (Figure 14, Figure 15) gases mediums. From 200 Pa pressure packaging the $S_{EM}$, at static condition, the argon gas presents the better stability compared with the remaining gases ($S_{EM} = 0.65$ pF/g) from temperature sweep range. On the other hand, $S_{EM}$ of helium gas varying around two percent for the same temperature range and held on frequency. The argon gas presented an intermediate performance assuming the same conditions. Such variations is due to the different properties of gas mediums, since the mechanical thermoelastic effect is the same for the transducer structure. Also, this observation corroborate with the fact of the Si to be a material with good thermal stability and appropriate for sensing applications [23]. When the frequency raising, with the pressure value equal to 200 Pa, the $S_{EM}$ values decreasing substantially, for all gases mediums. The simulation results obtained using atmosphere packaging at 20 Pa, shows an improvement in $S_{EM}$ variations at static condition and also in frequency linearity response.

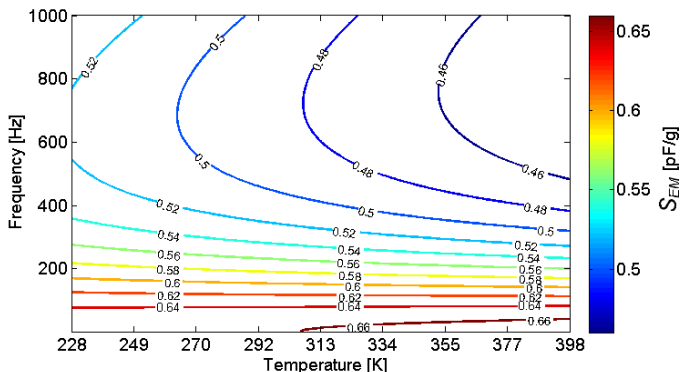

Figure 10 – Dynamic electro-mechanical sensibility simulation with temperature range from 228 K to 398 K and frequency range from 0 to 1000 Hz, for air at 200 Pa atmosphere packaging.

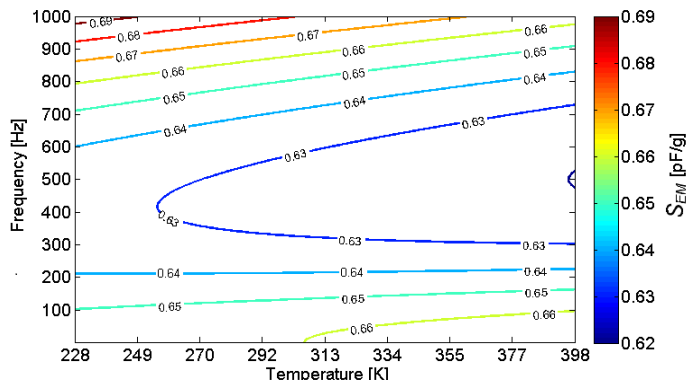

Figure 11 – Dynamic electro-mechanical sensibility simulation with temperature range from 228 K to 398 K and frequency range from 0 to 1000 Hz, for air at 20 Pa atmosphere packaging.

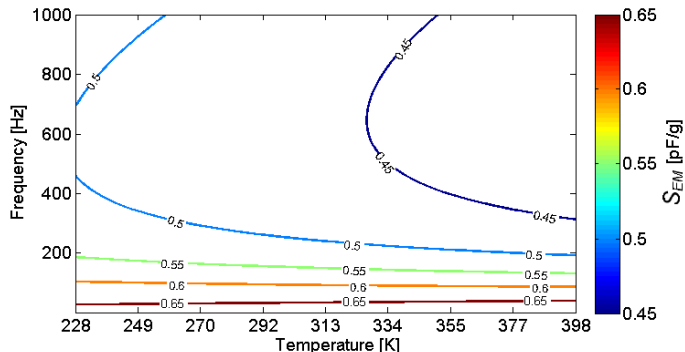

Figure 12 – Dynamic electro-mechanical sensibility simulation with temperature range from 228 K to 398 K and frequency range from 0 to 1000 Hz, for argon at 200 Pa atmosphere packaging.

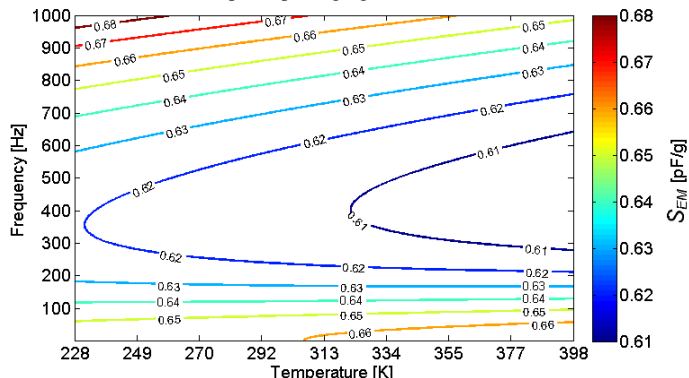

Figure 13 – Dynamic electro-mechanical sensibility simulation with temperature range from 228 K to 398 K and frequency range from 0 to 1000 Hz, for argon at 20 Pa atmosphere packaging.

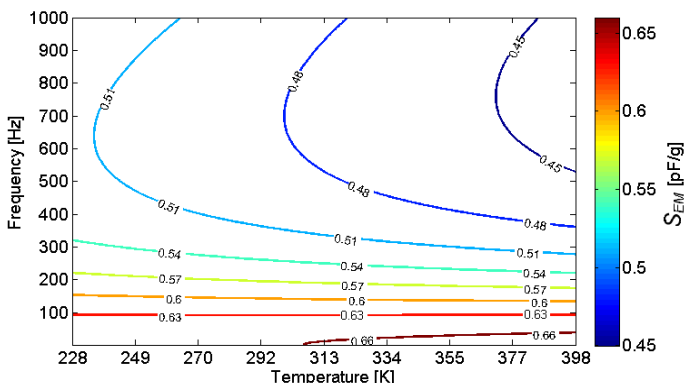

Figure 14 – Dynamic electro-mechanical sensibility simulation with temperature range from 228 K to 398 K and frequency range from 0 to 1000 Hz, for helium at 200 Pa atmosphere packaging.

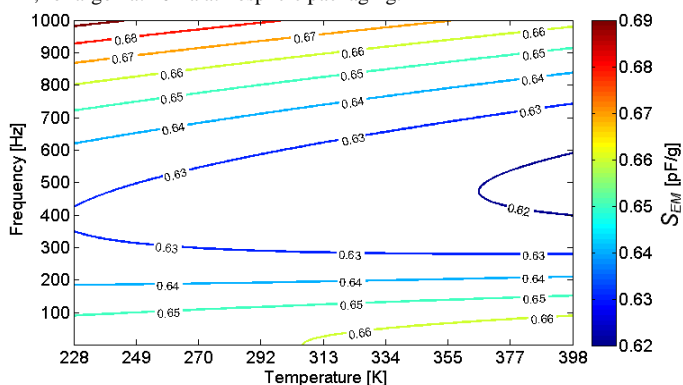

Figure 15 – Dynamic electro-mechanical sensibility simulation with temperature range from 228 K to 398 K and frequency range from 0 to 1000 Hz, for helium gas at 20 Pa atmosphere packaging.

The results show that is important to choose the most inert gas medium, if the MEMS transducer packaging process is not carried out at high vacuum conditions. In the case of the present study, the best gas to use is helium. But, if high vacuum conditions are adopted during the MEMS transducer package process, the gas medium is not a determinant factor to establish the accuracy and requisites of the sensing device.

## 4. Conclusion

The theoretical dynamic behavior of capacitive MEMS acceleration transducers based on silicon buk-micromachined design was presented and evaluated. This adopted analytical model considers the squeeze film damping effects for the molecular regime and the mechanical thermoelasticy in order to represent the electro-mechanical sensitivity in the frequency domain. The thermal stability of silicon, and of the other materials considered, results variations of $S_{EM}$ values around two percent at the proposed temperature range. In addition, the main damping loss mechanism related to the SFD effects were presented. The internal atmosphere , composed either by air, argon, or helium, was simulated, using the proposed model approaches. As a simulation results, using a small values of pressure inside the MEMS transducer was obtained the best linearity responses. The good choice of the gas medium in MEMS package is dependent of the capacity to provide a high vacuum inside the transducer, and so it will determine the accuracy and exigency of the application.

## Acknowledgment

This work was funded by FINEP, n: 01.09.0395.00, and CNPq, n: 559908 / 2010-5. A.P and V.R.A. thank CNPq, n: 310578 / 2012-4 and 312029 / 2013-6, respectively. V.R.A. thanks CAPES for PVS-CAPES/ITA n. 48/2014. L.L.S. and J.R.R thank CAPES for the scholarship and all ACELERAD project stakeholders for their support in preparing this paper.